\documentclass{aa}  
\usepackage{graphicx}
\usepackage{multirow}
\usepackage{xcolor}
\newcommand{\chandra}{{\sl Chandra}~}
\newcommand{\xmm}{{\sl XMM-Newton}~}
\newcommand{\spitzer}{{\sl Spitzer}~}
\usepackage{graphicx}

\usepackage[normalem]{ulem}

\usepackage[varg]{txfonts}
\begin{document}

   \title{The discovery of radio halos in the Frontier Fields clusters Abell\,S1063 and  Abell\,370}


   \author{C.~Xie
          \inst{1}
          \and
           R.~J.~van~Weeren\inst{1}
           \and
            L.~Lovisari\inst{2}
          \and
          F.~Andrade-Santos\inst{2}
          \and
          A.~Botteon\inst{1,3}
          \and
          M.~Br\"uggen\inst{4}
          \and
          E.~Bulbul\inst{2}
         \and
          E.~Churazov\inst{5,6}
          \and
          T~.E.~Clarke\inst{7}
          \and
          W.~R.~Forman\inst{2}
          \and
          H.~T.~Intema\inst{8,1}
          \and 
          C.~Jones\inst{2}
          \and
          R.~P. Kraft\inst{2}
         \and
          D.~V.~Lal\inst{9}
          \and
          T.~Mroczkowski\inst{10}
          \and
          A.~Zitrin\inst{11}
          }

   \institute{Leiden Observatory, Leiden University, PO Box 9513, 2300 RA Leiden, The Netherlands\\ \email{xie@strw.leidenuniv.nl, rvweeren@strw.leidenuniv.nl}
        \and
        Center for Astrophysics | Harvard \& Smithsonian, 60 Garden Street, Cambridge, MA 02138, USA
        \and
       INAF - IRA, via P.~Gobetti 101, I-40129 Bologna, Italy 
       \and
       Hamburger Sternwarte, University of Hamburg, Gojenbergsweg 112, 21029 Hamburg, Germany
       \and      
       Max Planck Institute for Astrophysics, Karl-Schwarzschild-Str 1, D-85741 Garching, Germany 
       \and
       Space Research Institute (IKI), Profsoyuznaya 84/32, Moscow 117997, Russia
       \and
       Naval Research Laboratory, 4555 Overlook Avenue SW, Code 7213, Washington, DC, 20375, USA
       \and
       International Centre for Radio Astronomy Research -- Curtin University, GPO Box U1987, Perth, WA 6845, Australia
       \and
       National Centre for Radio Astrophysics - Tata Institute of Fundamental Research, Post Box 3, Ganeshkhind P.O., Pune 411007, India
       \and 
       European Southern Observatory (ESO), Karl-Schwarzschild-Str. 2, D-85748 Garching, Germany
       \and
       Physics Department, Ben-Gurion University of the Negev, P.O. Box 653, Beer-Sheva 8410501, Israel
       \\
       \\
            }

   \date{Received XXX; accepted XXX}

 
  \abstract
   { Massive merging galaxy clusters often host diffuse Mpc-scale radio synchrotron emission. This emission originates from relativistic electrons in the ionized intracluster medium (ICM). An important question is
how these synchrotron emitting relativistic electrons are accelerated. 
 }
   {Our aim is to search for diffuse emission in the Frontier Fields clusters Abell S1063 and Abell 370 and characterize its properties. While these clusters are very massive and well studied at some other wavelengths, no diffuse emission has been reported for these clusters so far.
   }
   {We obtained 325~MHz Giant Metrewave Radio Telescope (GMRT) and 1--4~GHz Jansky Very Large Array (VLA) observations of Abell\,S1063 and Abell\,370. We complement these data with \chandra and \xmm X-ray observations.
   }
   {In our sensitive images, we discover radio halos in both clusters. In Abell\,S1063, a giant radio halo is found with a size of $\sim 1.2$~Mpc. The integrated spectral index between 325~MHz and 1.5~GHz is $-0.94\pm0.08$ and it steepens to $-1.77 \pm 0.20$ between 1.5 and 3.0~GHz.
   This spectral steepening provides  support for the turbulent re-acceleration model for radio halo formation.
    Abell\,370 hosts a faint radio halo mostly centred on the southern part of this binary merging cluster, with a size of $\sim 500-700$~kpc. The spectral index between 325 MHz and 1.5 GHz is $-1.10\pm0.09$. Both radio halos follow the known scaling relation between the cluster mass proxy $Y_{500}$ and radio power, consistent with the idea that they are related to ongoing cluster merger events.
   }
   {}
   


\titlerunning{The discovery of radio halos in Abell S1063 and Abell 370}
\maketitle
%

\section{Introduction}

Diffuse radio sources in galaxy clusters trace large-scale magnetic fields and relativistic electrons in the intracluster medium (ICM). Unlike the synchrotron emission from radio galaxies, diffuse cluster radio sources do not directly associate with any individual sources in the cluster. Diffuse radio sources are commonly divided into three different classes: radio halos, radio mini-halos, and radio relics/shocks (see, e.g. \citealp{Feretti2012}, \citealp{Brunetti2014}, and \citealp{vanWeeren2019} for reviews). Radio shocks (relics) are elongated, arc-like objects located in the periphery of  merging clusters \citep[e.g.,][]{2007A&A...463..937V, 2009A&A...494..429B, 2009A&A...506.1083V, 2016ApJ...818..204V}. On the other hand, radio halos and mini-halos are located at the cluster center with more roundish morphologies \citep[e.g.,][]{ 1993ApJ...406..399G, 1998A&A...331..901S, 2011MNRAS.412....2B, vanWeeren2014, Kale2015, Cuciti2018, Giacintucci2019}.




Radio halos have typical sizes of $\sim1$~Mpc. The spectral indices\footnote{The spectral index $\alpha$ is defined as $F_{v} \propto v^{ \alpha} $.} of radio halos are steep, ranging from about $-1$ to $-2$. Radio halos approximately follow the X-ray emission from the hot ICM  \citep[e.g.,][]{2004ApJ...605..695G}, indicating a connection between the thermal and non-thermal components of the ICM. This connection also supported by the correlation between the radio power and X-ray luminosity  or cluster mass \citep[e.g.,][]{2000ApJ...544..686L,2006MNRAS.369.1577C,Basu2012, Cassano2013}. So far, most Mpc-size radio halos have been found in dynamically disturbed clusters, suggesting a connection between mergers and radio halo formation \citep[e.g.,][]{2010ApJ...721L..82C}.


There are two main models to explain the origin of radio halos. The turbulent re-acceleration model proposes that the cosmic-ray (CR) electrons are re-accelerated by the turbulence induced by a cluster merging event \citep{Brunetti2001, Petrosian2001}. In the hadronic model for radio halos, the relativistic electrons are secondary products produced by proton-proton collisions \citep{Dennison1980, Blasi1999}. Such collisions will also produce $\gamma$-ray emission. Recent Fermi-LAT observations gave important constraints on the energy content of CR protons in clusters \citep{Ackermann2014_Fermi_gammaray, Ackermann2016_Fermi_gammaray, Brunetti2017}, which disfavour the hadronic model. 
The discovery of ultra-steep spectrum radio halos \citep[i.e., $\alpha$~\textless $-1.6$;][]{Brunetti2008} has also provided support for the turbulent re-acceleration model. Despite these findings, our understanding of the turbulent re-acceleration mechanism remains limited.


Radio mini-halos have typical sizes of $\lesssim500$~kpc \citep[e.g.,][]{Feretti2012,2017ApJ...841...71G}. The most prominent difference with giant radio halos is that mini-halos are not associated with merging clusters but with relaxed, cool-core clusters. Such clusters often contain a radio-loud active galactic nucleus (AGN) at their center. However, the radiative lifetime of the CR electrons from radio mini-halos is too short for these electrons to have directly come from the central AGN. Therefore in-situ particle (re-)acceleration in the ICM is required to explain the existence of mini-halos.


Radio mini-halos have been explained by turbulent re-acceleration from gas sloshing in the cluster core \citep{2008ApJ...675L...9M,2013ApJ...762...78Z}. However, hardonic scenarios have also been proposed \citep[e.g.,][]{2004A&A...413...17P,2007ApJ...663L..61F,2010ApJ...722..737K,2013MNRAS.428..599F}. Although mini-halos are sometimes considered as smaller versions of giant halos, the connection between halos and mini-halo is still unclear \citep{Savini2019,vanWeeren2019,2019MNRAS.486L..80K}. 



\begin{table*}[th!]
\caption{Clusters properties}             
\label{table:1}      
\centering                          
\begin{tabular}{c c c c c c}        
\hline\hline                 
Name & RA & Dec & Redshift & $M_{\rm{SZ}}^{a}$  & $ Y_{500}^{b}$  \\    
  & (J2000) & (J2000) &  & (10$^{14}$ M$_{\odot}$) & ($10^{-4} \rm{Mpc}^{2}$)\\
\hline    
   Abell S1063 & 22 48 43.5 & -44 31 44.0 & 0.346  &$11.4 \pm 0.3$ &    $2.32 \pm 0.23$  \\
   Abell 370   & 02 39 50.5 & -01 35 08.2 & 0.375  &$7.6 \pm 0.6$ &   $1.75 \pm 0.46$  \\      

\hline
\end{tabular}
\tablefoot{$^{(a)}$~From Planck measurements \citep{planck15}; $^{(b)}$~The Y$_{5R_{500}}$ from Planck measurements \citep{planck15} was rescaled to Y$_{500}$ = Y$_{5R_{500}}$ / 1.79 \citep{Arnaud2010}.}
\end{table*}

In this paper, we present new 325~MHz Giant Metrewave Radio Telescope (GMRT) and 1--4~GHz Jansky Very Large Array (VLA) observations of two Frontier Fields clusters, \object{Abell\,S1063} and \object{Abell\,370}. Deep VLA observations of the Frontier Fields cluster MACS~J0717.5+3745 and Abell\,2744 were already presented in \cite{2016ApJ...817...98V,2017ApJ...835..197V} and \cite{2017ApJ...845...81P}. The observations and data reduction are described in Section~\ref{sec:obsdata}. In Section~\ref{sec:results}, we present our results and radio spectral measurements. We end with a discussion and conclusions in
Sections~\ref{sec:discussion} and~\ref{sec:conclusions}. 
Below we introduce these clusters in some more detail. 


Throughout the paper, we adopt the flat $\Lambda$CM cosmology with $\Omega_{\Lambda} = 0.70$, $\Omega_{M} = 0.30$, and H$_{0}$ = 70 km s$^{-1}$ Mpc$^{-1}$. At the redshifts of Abell 370 and  Abell S1063, 1\arcsec corresponds to scales of about 5.2 kpc and 4.9 kpc, respectively.

\subsection{Abell S1063 and Abell 370}
\label{sec:introclusters}


Abell S1063 (also known as RXC J2248.7--4431 or MACS 2248.7-4431, hereafter AS1063) is a massive galaxy cluster ($z=0.3461$) with a Sunyaev-Zel'dovich (SZ) derived mass of $M_{500} \sim 1.4 \times 10^{15} M_{\odot}$ \citep{Abell, planck15, Lotz}, see Table~\ref{table:1}. The cluster's X-ray luminosity is 1.8 $\times$ 10$^{45}$ erg s$^{-1}$ in the 0.5--2.0~keV band \citep{SPTsurvey} and the mean temperature is around 12~keV within a 800~kpc radius,
which makes it one of the hottest clusters known \citep{Gomez2012}.

An observed offset between the galaxy isodensity distribution
and hot gas, high X-ray temperature, and non-Gaussian galaxy velocity distribution suggest an ongoing major merger event  \citep{Gomez2012}. A subsequent weak lensing study provided further support for the conclusion of an ongoing merger \citep{Gruen2013}. However, from \xmm observations, the high concentration value, low power ratio, and low centroid shift of the X-ray emission lead \cite{Lovisari2017} to suggest a relaxed dynamical state. It should be noted that these parameters may lead to wrong conclusions if the mergers happen along the line of sight or with a small offset.
 

Abell 370 (hereafter, A370) is renown for being a strong lensing cluster ($z=0.375$) with a mass of $M_{500} \sim 1.1 \times 10^{15} M_{\odot}$ \citep{Abell, Struble1999, Morandi2007, planck15, Lotz}. The bolometric X-ray luminosity is 1.1~$\times$~10$^{45}$ erg~s$^{-1}$ \citep{Morandi2007}. A370 is the first cluster that was found to gravitationally lens a background galaxy \citep{Hoag1981, Lynds1986, Soucail1987, Paczynski1987}. 
Since then, numerous works studied its properties \citep{Kneib1993, 1996ApJ...469..508S, Broadhurst2008, Richard2010, Umetsu2011, Lagattuta2017, Strait2018}. The  matter distribution of the cluster shows two main substructures, one centered on the northern and one on the southern brightest cluster galaxy (BCG),  indicating a recent merging event \citep{Richard2010}. The velocity dispersion of each subcluster is about 850~km~s$^{-1}$ \citep{Kneib1993}. The dynamical unrelaxed state of the cluster is also suggested by the presence of X-ray surface brightness edges in the ICM, that may be related to shocks and/or cold fronts \citep{Botteon2018}.

Previous studies of A370 did not report on any extended radio emission in the cluster. \cite{Lah2009} analyzed the hydrogen gas content of 324 galaxies around the cluster based on GMRT observations. \cite{Wold2012} catalogued the radio sources in the cluster field, with VLA observation in A and B configurations. Some radio galaxies in the cluster field have also been studied \citep{Smail2000, Hart2009}. 



\section{Observations and data reduction}
\label{sec:obsdata}
\subsection{VLA observations}

AS1063 and A370 were observed by the VLA in L- and S-bands with multiple array configurations (project: 16B-251, PI: R.J. van Weeren). The details of radio observations can be found in Table~\ref{logofobs}.  Due to the low declination of AS1063, no D-array observations were obtained. The total on-source time for AS1063 and A370 are 13~hrs and 9~hrs, respectively. Recorded bandwidths are 1 GHz (L-band) and 2 GHz (S-band), covered by 16 spectral windows, with 64 channels each. The primary calibrators we used are 3C138 and 3C147. For the phase calibrator, we used J2214-3835 and J0149+0555 for AS1063 and A370, respectively.

The data were calibrated and reduced using the Common Astronomy Software Applications ({\tt CASA}; McMullin et al. 2007) package, version 5.1.1. The data observed in different runs were processed separately using the same procedures. Each data set was first Hanning smoothed and data affected by antenna shadowing were flagged. Radio frequency interference (RFI) was automatically flagged using the {\tt CASA} `tfcrop' mode in the {\tt flagdata} task. Manual flagging was also applied if the antenna was not working properly by inspecting the bandpass, gain, and polarisation solutions. The elevation dependent gain tables and antenna offsets positions were also applied. After that, the initial gain solutions of primary calibrators 3C147 and 3C138 were determined based on the central ten channels of a spectral window to remove the phase variation during calibrator observations. These initial gain solutions were applied to find the antenna-based delays 
and bandpass calibration tables. Applying the delay and bandpass solutions, we re-determined the gain solutions for our primary calibrators using the full bandwidth, apart from a few noisy edge channels. 
Next, we determined the gain solutions for phase calibrators. The flux scale of primary calibrators was calculated from the \cite{Perley&Butler2013} model and applied to the phase calibrators.
As a final step, we applied all the solutions to the target field and averaged the data by a factor of 3 and 4 in time and frequency, respectively. To remove additional RFI, we performed a last flagging step with {\tt AOFlagger} \citep{AOFlagger} on the target field data.

To refine the calibration of the target field, two rounds of phase-only self-calibration were applied. This was followed by a few amplitude and phase self-calibration rounds until the image quality did not improve further. Additional bad data were flagged during the self-calibration by visually inspecting the gain solutions. The imaging during the self-calibration was done with {\tt CASA}, using the full bandwidth to make a deep Stokes~I image with Briggs weighting \citep[robust=0;][]{Briggs}. To account for the non-coplanarity of the array, w-projection with 256 planes \citep{W-projection} was employed. The clean masks were created by the Python Blob Detector and Source Finder ({\tt PyBDSF}; \cite{PyBDSM}) package. For the wide-band deconvolution, the spectral index was taken into account with `nterms' of 3 \citep{nterms}. After the self-calibration, we combined the data sets from the different array configurations in the same frequency band and run an extra self-calibration step to align them. The primary beam attenuation was also corrected for.

\subsection{GMRT observations}

AS1063 and A370 were observed with GMRT at 325 MHz with a bandwidth of 33.3 MHz (project code: 31\_037, PI: R.J. van Weeren) on different observing sessions (see, Table~\ref{logofobs}). We used the Source Peeling and Atmospheric Modeling ({\tt SPAM}; \cite{spam}) pipeline to process the continuum observations obtained with the GMRT software correlator backend (GSB). The details of SPAM pipeline can be found in \cite{Intema2014, Intema2017}. We combined the multiple data sets for the single targets  during the {\tt SPAM} processing. To summarise, the {\tt SPAM} pipeline performs direction-independent and direction-dependent calibration. The main steps include the averaging and flagging of data, bandpass and flux scale calibration \citep{Scaife&Heald2012}, initial phase-only calibration, and direction dependent calibration and ionospheric modeling using the bright sources in the primary beam.


The output from SPAM was imaged using {\tt CASA} with w-projection (256 planes) and Briggs weighting, robust 0. The details of imaging parameters can be found in Table~\ref{imaging information}. For the GMRT data, `nterms' of 2 was adopted.

\subsection{Flux density uncertainties and compact source subtraction}
\label{sec:sourcesubtraction}
Throughout the paper, the uncertainties on the flux density measurements are estimated using
\begin{equation}
\sigma = \sqrt{\sigma_{\rm cal}^{2} + \sigma_{\rm R}^{2}},
\end{equation}
where the statistical error $\sigma_{\rm R} = \sigma_{\rm rms} \times \sqrt{N_{\rm beams}}$, with the noise level of image $\sigma_{\rm rms}$ and the number of beams $N_{\rm beams}$ covered by the source. The absolute flux-scale calibration uncertainty is $\sigma_{\rm cal} = fS_{\rm{int}}$, with $S_{\rm{int}}$ the flux density and $f$ the fractional uncertainty of the flux-scale. We adopt $f=0.1$ for the GMRT and $f=0.05$ for the VLA. 


To search for and characterize the diffuse emission in the two clusters, the contribution from compact sources needs to be removed. We did this by first imaging the combined data sets of each frequency band using robust=-1 weighting and an inner uv-cut of 3~k$\lambda$. At the redshift of our clusters, 3~k$\lambda$ corresponds to about 400~kpc. 
We then computed the visibility data of this model for the entire uv-data range and subtracted these from the calibrated visibility data using the {\tt CASA} task {\tt uvsub}.


For imaging possible diffuse emission in the clusters, we used multiscale clean \citep{W-projection} with scales of [0, 3, 7, 25, 75]\footnote{The scales are in unit of pixels. The beam is sampled by $\sim$4 pixels.}, uvtapers, and Briggs weighting. The details of the imaging parameters can be found in Table~\ref{imaging information}. 
The integrated flux densities for diffuse emission, or upper limits, are computed from these images. Based on the residuals, at the location of bright compact sources outside the cluster region, we estimate the error on the compact subtraction is less than 1\%. To provide an alternative estimate for the integrated flux densities, we also measured the integrated flux densities using images that still contained the compact sources. The contribution of the compact sources was then removed by manually computing their integrated flux densities on our images with the highest spatial resolution (Table~\ref{imaging information}). In this case, we include the uncertainty on the subtraction of the compact sources, on the total uncertainty of the flux density of the diffuse emission, using standard error propagation. In all cases, the results using our two methods gave results that were consistent. For completeness, we report the flux densities using the latter method in Table~\ref{flux} with footnotes.

\subsection{Spectral index maps}

To map the spectral index distribution, we used an inner uv- cut of 0.15~k$\lambda$ for AS1063 and 0.22~k$\lambda$ for A370, to sample the same spatial scales at all three frequency bands. Note that for A370, due to the non-detection of diffuse emission in the S-band, we chose the shortest baseline only from on L-band and 325MHz data to maximally recover the diffuse flux. The {\tt CASA} tasks {\tt imsmooth} and {\tt imregrid} were used to align the beam shapes and pixel grids. Only pixel values larger than 3$\sigma_{rms}$ were used to calculate the spectral index maps. The integrated flux densities were also measured from the same images that were used to construct spectral index maps.

\subsection{X-ray observations}
\label{sec:xray}

Abell S1063 was observed with the {\em Chandra} X-ray Observatory
(ACIS-I detectors, VF mode, ObsIds 4966 -- PI Romer, 18611, 18818 -- PI Kraft). {Abell~370 was observed with ACIS-I and S detectors, ObsIds 515 and 7715 -- PI Garmire}. The data were reduced using the software \texttt{CHAV} which follows the processing described in
\citet{2005Vik}, applying the calibration files \texttt{CALDB 4.7.1}. 
The data processing includes corrections for the time dependence of the charge
transfer inefficiency and gain, and a check for periods
of high background.
Also, readout artifacts were
subtracted and standard blank sky background files were used for 
background subtraction. We show the combined images of all observations in the 0.5--2.0 keV energy band in Section~\ref{sec:results}.

\xmm observations were  processed with the XMMSAS v16.0.0 software\footnote{https://www.cosmos.esa.int/web/xmm-newton/sas}. The calibrated event files were generated from raw data by running the tasks {\it emchain} and {\it epchain}. Throughout this analysis, we only considered single pixel events for the pn data
(i.e., PATTERN==0) and single, double, triple, and quadruple events (i.e. PATTERN$\le$12) for MOS. In addition, we removed all the events next to CCD edges and next to bad pixels (i.e., FLAG==0) and we applied the pn out-of-time correction.
All the data sets were cleaned for periods of high background due to solar flares following the two stage filtering process extensively described in \cite{Lovisari2011}. The task {\it edetect-chain} has been used to detect the point-like sources which were then excluded from the event files. The background event files were cleaned by applying the same PATTERN selection, flare rejection criteria and point-source removal used for the observation events. 

The background-subtracted and exposure-corrected image for A370 (see Fig. 8) has been obtained with MOS data only in the 0.3-7 keV band using a binning of 40 physical pixels corresponding to a resolution of 2 arcsec, and smoothed with a Gaussian of FWHM of 6 arcsec. The background subtraction was performed using a combination of blank-sky-field and filter-wheel-closed observations as described in \cite{Lovisari2017}.

The temperature profile of AS1063 has been derived using successive annular regions centred on the X-ray peak. The size of the annuli was determined by requiring that the width is larger than 0.5$^{\prime}$, to minimize the flux redistribution due  to the PSF, and a S/N$>$100 to measure the temperatures with an accuracy of $\sim10\%$ (68\% confidence limit). The spectral fitting procedure and the modeling of the different background components is fully described in \cite{Lovisari2019}.

\begin{table*}

\caption{Log of radio observations}             
\label{logofobs} 
\centering

\begin{tabular}{lcccccc}
\hline\hline
Name & Observation &  Observing date 	& Frequency coverage 	& Channel width & Integration time	& On-source time	\\
	 &				& 			    & (GHz)		     	& (MHz)		    & (s)				& (hr)				\\

\hline
\multirow{10}{*}{\textbf{Abell S1063}}	
&\multirow{2}{*}{GMRT 325 MHz}	    & 25 Feb 2017	 & 0.31--0.34	& 0.13		& 16			& 4.8       \\
&		                            & 26 Aug 2017    & 0.31--0.34    & 0.13		& 4				& 4.1		 \\
\noalign{\smallskip}
&\multirow{2}{*}{ L-band C-array}	& 24 Jun 2017		 & 1--2		& 1			& 5				& 2.0		\\
&		                            & 25 Jun 2017        & 1--2	    & 1			& 5				& 2.0		\\
&\multirow{2}{*}{ L-band B-array}	& 13 Oct 2017		 & 1--2		& 1			& 3				& 1.3		\\
&		                            & 07 Nov 2017        & 1--2	    & 1			& 3				& 1.3		\\
\noalign{\smallskip}
&\multirow{2}{*}{ S-band C-array}	& 17 Jun 2017		 & 2--4		& 2			& 5				& 1.9		\\
&		                            & 18 Jun 2017        & 2--4	    & 2			& 5				& 1.9		\\
&\multirow{2}{*}{ S-band B-array}	& 02 Oct 2017		 & 2--4		& 2			& 3				& 1.2		\\
&		                            & 04 Oct 2017        & 2--4	    & 2			& 3				& 1.2		\\
\noalign{\smallskip}

\hline
\multirow{7}{*}{\textbf{Abell 370}}
&\multirow{3}{*}{GMRT 325 MHz}   	& 07 Jan 2017	 & 0.31--0.34	& 0.13		& 8				& 4.8			\\
&		                            & 05 Mar 2017    & 0.31--0.34	& 0.13		& 4				& 5.7			\\
&		                            & 01 Aug 2017    & 0.31--0.34	& 0.13		& 4				& 4.7			\\
\noalign{\smallskip}
&L-band D array                 & 11 Feb 2017	    & 1--2		& 1			& 5				& 1.5		        \\	
&L-band C array	                & 20 May 2017	    & 1--2		& 1			& 5				& 3.2				\\
\noalign{\smallskip}
&S-band D array	                & 11 Feb 2017	    & 2--4		& 2			& 5				& 1.3				\\
&S-band C array	                & 28 May 2017 	    & 2--4		& 2			& 5				& 3.1				\\
\noalign{\smallskip}

\hline
\end{tabular}

\end{table*}


\begin{table}
\caption{Imaging information}             
\label{imaging information}      
\centering                          
\begin{tabular}{lccccc}        
\hline\hline                 
AS1063  &  uv-cut & uvtaper & beam & $\sigma_{\rm rms}$  \\    
      &  (k$\lambda$)        & (\arcsec)& (\arcsec $\times$ \arcsec) & ($\mu$Jy~beam$^{-1}$)\\
\hline                        

   325MHz   & --      & --    & 26.0 $\times$ 9.0  &  46 \\
   325MHz   & \textgreater 0.15  & 15   & 45.0 $\times$ 16.0  & 96 \\
   L-band   &  --     & --    & 26.0 $\times$ 9.0  & 21  \\      
   L-band   & \textgreater 0.15  & 15   & 45.0 $\times$ 16.0  & 40 \\      
   S-band   & --      & --    & 26.0 $\times$ 9.0  & 11 \\      
   S-band   & \textgreater 0.15  & 15   & 45.0 $\times$ 16.0  & 24 \\ 
\hline\hline
A370  &  uv-cut & Uvtapers & Beam & $\sigma_{\rm rms}$  \\    
      &  (k$\lambda$)        & (\arcsec)& (\arcsec $\times$ \arcsec) & ($\mu$Jy~beam$^{-1}$)\\
\hline
   325MHz   & --  & --& 11.9 $\times$ 8.6  & 62 \\
   325MHz   & \textgreater 0.22  & 20& 25.0 $\times$ 22.0  & 161 \\

   L-band   & --  & --& 14.8 $\times$ 11.5  & 26  \\      
   L-band   & \textgreater 0.22   & 20 & 25.0 $\times$ 22.0  & 34  \\      
   S-band   &  -- & --& 8.5 $\times$ 7.3  & 9   \\      
   S-band   &  -- & 20 & 25.0 $\times$ 22.0  & 18 \\

\hline
\end{tabular}
\tablefoot{All images are made with Briggs weighting, robust 0.}

\end{table}

%
%
   \begin{figure}
   \centering
    \includegraphics[width=0.49\textwidth]{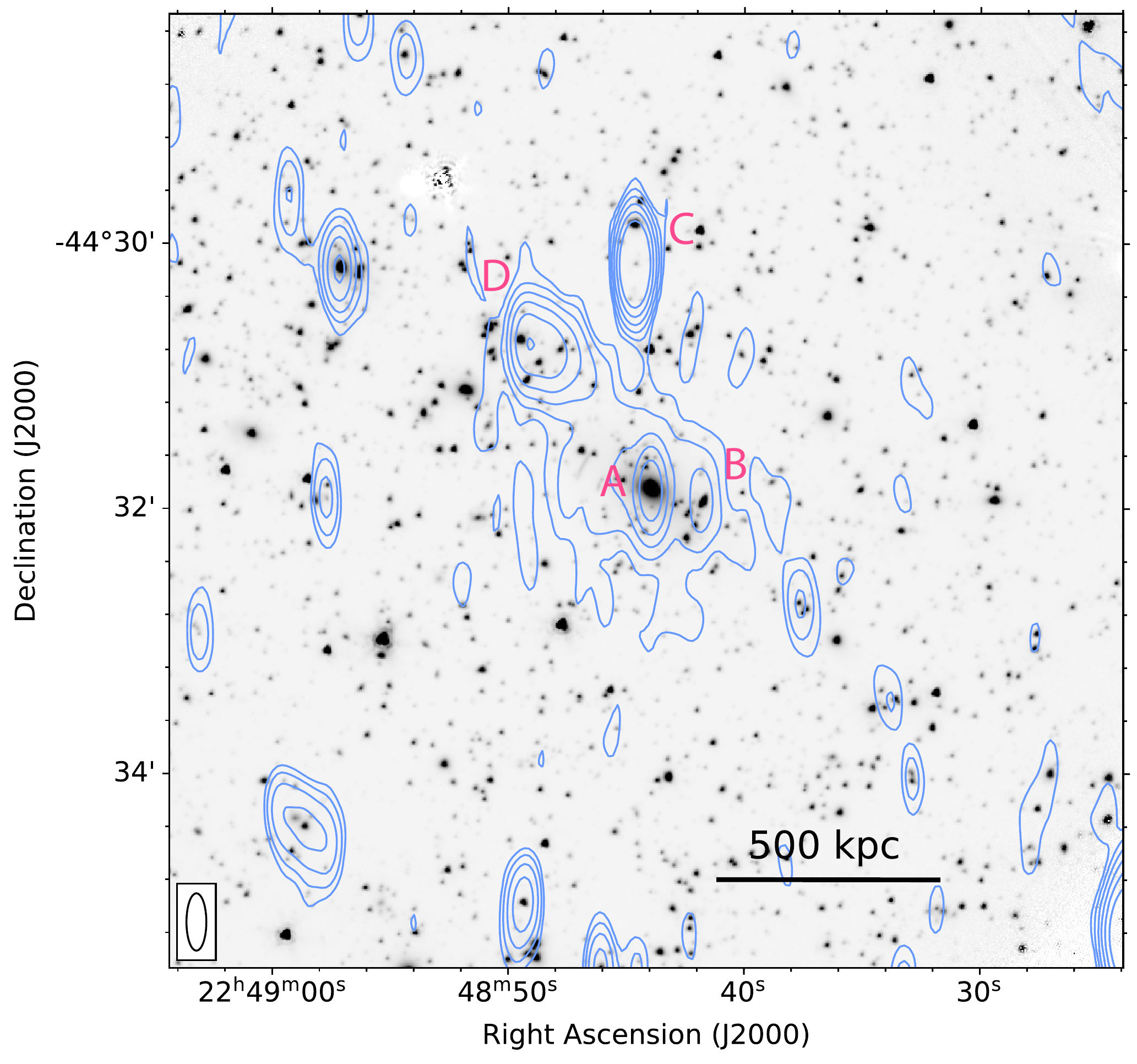}
     \caption{{\sl Spitzer} 3.6~$\mu$m IRAC image of AS1063 overlaid with VLA 1.5~GHz radio contours. Contour levels are drawn at [1, 2, 4, 8, 16, 32] $\times$ 3$\sigma_{\rm rms}$, where $\sigma_{\rm rms}$ = 21 $\mu$Jy~beam$^{-1}$. Due to the low declination of the cluster, the radio beam is quite elongated (26\arcsec $\times$ 9\arcsec). Compact radio sources are labelled A to D, see also Figure~\ref{spec_index_map_SourceD}. }
    \label{Lband_IR_as1063}
   \end{figure}
%

\section{Results}
\label{sec:results}

\subsection{Abell\,S1063}

A {\sl Spitzer} infrared (IR) image of AS1063, overlaid with our 1.5~GHz radio contours is presented in Figure~\ref{Lband_IR_as1063}. Four compact radio sources are detected in the cluster vicinity, labelled A to D in Figure~\ref{Lband_IR_as1063}. In the central regions of the cluster more extended diffuse emission is also detected. The  properties of the radio sources are listed in Table~\ref{flux}.

%
Source~A, namely rxj2248\_17936\footnote{The ID number in CLASH \spitzer catalog \citep{Postman2012CLASH}.} or 2MASX J22484405-4431507, is the BCG in AS1063. The integrated flux densities of the BCG are $8.5\pm1.1$~mJy at 325~MHz, \textbf{$2.1\pm0.2$}~mJy at 1.5~GHz, and $1.0\pm 0.1$~mJy at 3.0~GHz. This corresponds to a spectral index of about $-1$, typical for a cluster AGN. Source~B (rxj2246\_18112) is a background galaxy at redshift of 0.61, identified as [GVR2012] 878 by \cite{Gomez2012}. Source~C (rxj2248\_19890) is the brightest radio galaxy in the cluster field, located north of the cluster center. 

Source~D (rxj2248\_18479) is a radio galaxy with a radio tail at the northeast of the cluster. The radio tail is connected to the diffuse emission and the tail length is $\sim$140 kpc at 1.5~GHz (see, Figure~\ref{Lband_IR_as1063}). The tail length increase at  325~MHz to $\sim$340~kpc. The integrated flux densities are $37.8\pm3.8$~mJy at 325~MHz, $5.6\pm0.3$~mJy at 1.5~GHz, and $1.8\pm0.1$~mJy at 3.0~GHz. The corresponding spectral indices are $\alpha^{1500}_{325}$ = -1.24 $\pm$ 0.07 and $\alpha^{3000}_{1500}$ = -1.61 $\pm$ 0.11. Such spectral behaviour indicates a high frequency break, the result of the radiative losses of the synchrotron emitting electrons.

   \begin{figure*}
   \centering
    \includegraphics[width=0.99\textwidth]{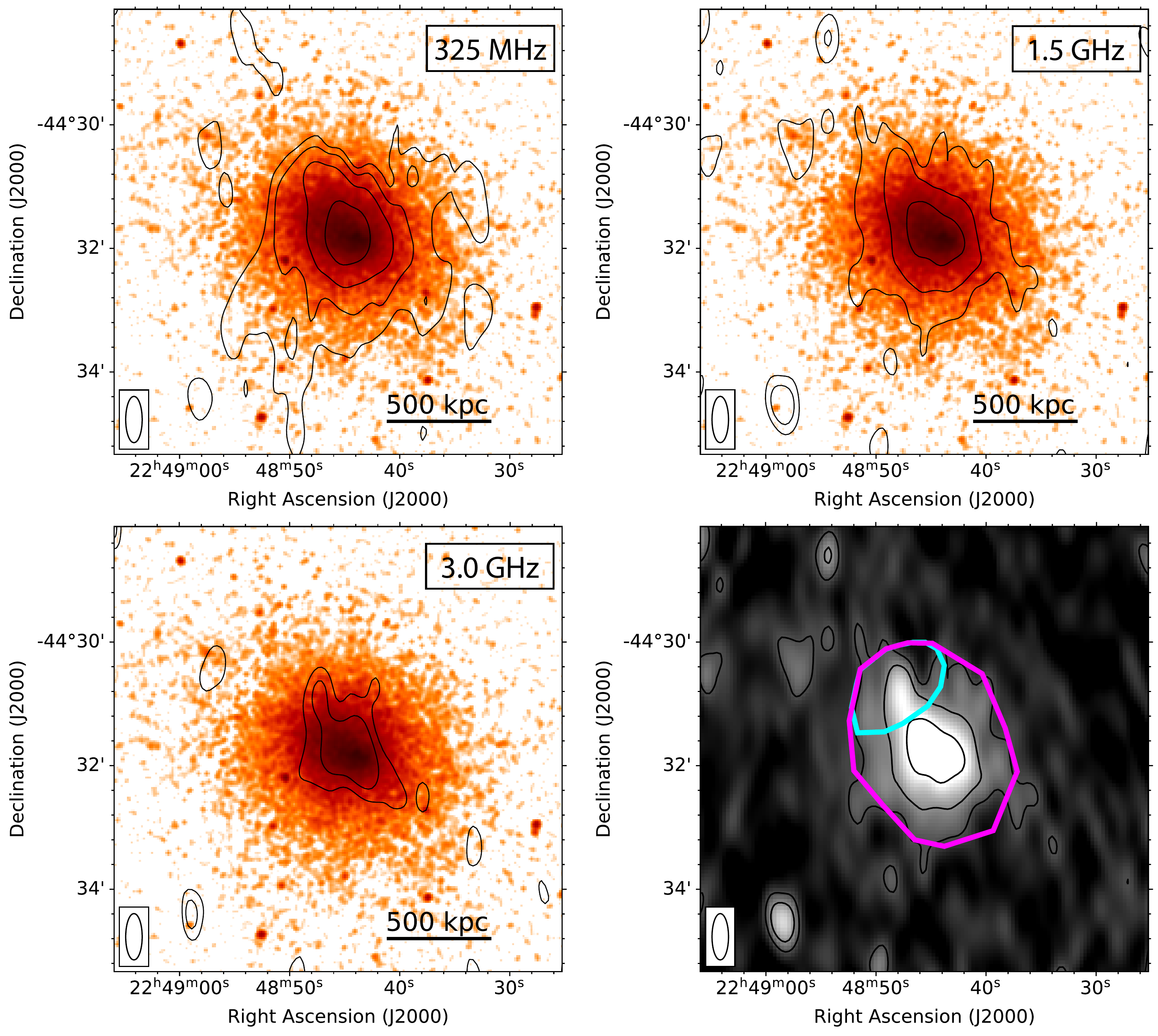}
      \caption{\chandra image of AS1063 overlaid with the radio contours of diffuse emission in AS1063 at three frequencies (325~MHz, top left; 1.5~GHz, top right; 3.0~GHz, bottom left). \chandra images are all in the energy band 0.5 -- 2.0 keV with pixel size 4 $\times$ 0.492\arcsec and smoothed with a Gaussian with scale of 3 pixels across.  Contour levels are drawn at [1, 2, 4, 8, ...] $\times$ 3$\sigma_{\rm rms}$, where  $\sigma_{\rm 325~MHz}$= 96 $\mu$Jy~beam$^{-1}$, $\sigma_{\rm 1.5~GHz}$= 40 $\mu$Jy~beam$^{-1}$, $\sigma_{\rm 3.0~GHz}$ = 24 $\mu$Jy~beam$^{-1}$.
      Compact radio sources were subtracted in all radio images. The inner uv-cut of 0.15~k$\lambda$ is adopted for all radio images.
      Bottom right: VLA L-band image of AS1063 depicting the region where we extract the integrated flux densities. The cyan polygon indicates the region where the diffuse AGN component is subtracted from the total flux measurement (magenta polygon).}
         \label{radio_X-ray}
   \end{figure*}

\begin{figure}
   \centering
    \includegraphics[width=0.48\textwidth]{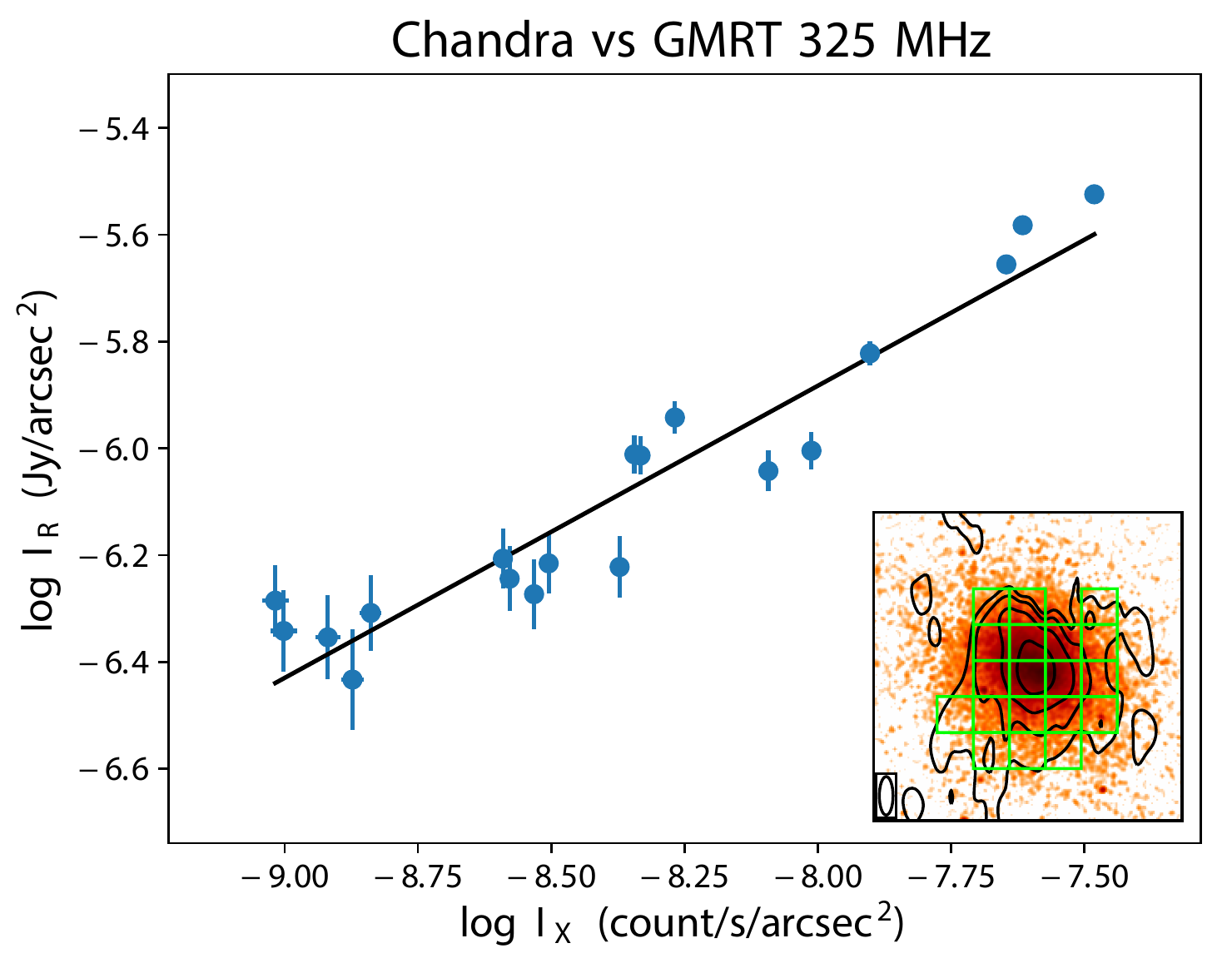}
      \caption{X-ray versus radio surface brightness (SB) for AS1063. The red line shows the best fit power-law $I_{\rm radio} \propto I_{\rm X-ray}^{ 0.55\pm0.04}$. The inset shows the region (green grid) where we measured the corresponding surface brightness.}
         \label{sb_vs_sb}
   \end{figure}

In Figure~\ref{radio_X-ray} we show the \chandra 0.5--2.0~keV X-ray image of the cluster. Radio contours at 0.325, 1.5 and 3.0~GHz are overlaid in the various panels.
The emission from compact sources was subtracted from the uv-data in these radio images (see Sect.~\ref{sec:sourcesubtraction}), to better determine the properties of the diffuse radio emission. Central diffuse emission is detected at all three frequencies which roughly follows the distribution of the thermal ICM. {We further investigate the connection between the thermal and non-thermal emission by comparing the X-ray and radio surface brightness evaluated in the same regions of the Chandra 0.5-2.0 keV and GMRT 325~MHz source subtracted images. Regions were chosen based on the $3\sigma$ level emission of the GMRT image, where 19 beam independent regions were identified. The plot of the radio versus X-ray surface brightness is reported in Figure~\ref{sb_vs_sb}, where regions with higher X-ray surface brightness seems associated to regions with higher radio surface brightness. This trend has been observed for a number of radio halos \citep{Govoni2001, Rajpurohit2018, Hoang2019, Cova2019}. Despite the small number of data points, we fit the data with a power-law in the form $I_{\rm radio} \propto I_{\rm X-ray}^b$ and obtain a slope b = 0.55 $\pm$ 0.04 which is within the range of values found in the literature. From the 1.5~GHz image, a slope of $0.48 \pm 0.07$ is obtained, which is consistent with the value from the 325~MHz image.} The largest physical extent of the diffuse emission is $\sim$700~kpc at 1.5~GHz and this increases to $\sim$1.2~Mpc at 325~MHz. Besides the central diffuse emission, we also detect remnant emission from the  tail of source~D (also see, Figure~\ref{Lband_IR_as1063}). Because of the extended nature of the tail, it is not possible to fully remove this in the source subtraction processes.  To determine the integrated flux density of the central diffuse emission we therefore exclude the area indicated by the cyan polygon in Figure~\ref{radio_X-ray} (bottom right panel). We used the same extraction region for the integrated flux density for all three frequencies (see Figure.~\ref{radio_X-ray}). The uncertainties are computed as described in Sect.~\ref{sec:sourcesubtraction}.

\begin{figure}
   \centering
    \includegraphics[width=0.51\textwidth]{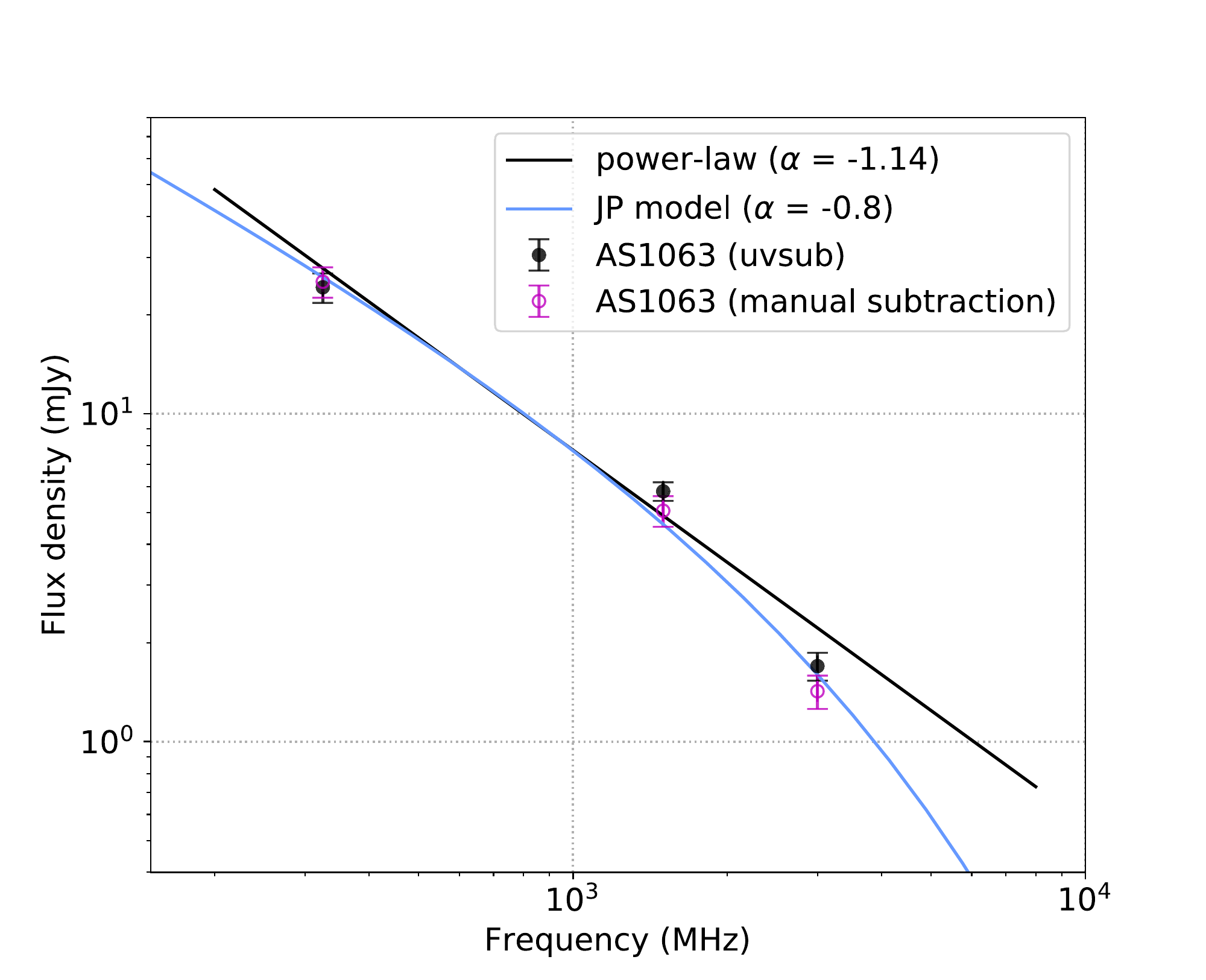}
      \caption{Integrated flux densities for the halo in AS1063 at 325~MHz, 1.5~GHz, and 3.0~GHz, measured by two methods. The black line shows a single power-law fit with spectral index $\alpha = -1.14$. The JP model has an injection spectral index of $-0.8$.  }
         \label{spec_index_as1063}
   \end{figure}

The flux densities of the central diffuse emission are $24.3 \pm 2.5$~mJy, $5.8 \pm 0.4$~mJy, and $1.7 \pm 0.2$~mJy, at 325MHz, L-band, and S-band, respectively. Using the observed flux densities, we find that the spectral index steepens at high frequencies with $\alpha^{1500}_{325} = -0.94~\pm~0.08$, and $\alpha^{3000}_{1500} = -1.77~\pm~0.20$. If we adjust the size of the extraction region to the extension of the radio halo at 325~MHz, the derived spectral indices remain consistent with each other within the uncertainties. In Figure~\ref{spec_index_as1063}, we present the integrated spectrum of the diffuse emission. A single power-law fit is unacceptable with $\chi^2$/d.o.f.~=17.7, showing the spectrum deviates from a power-law shape.

We conclude that the diffuse emission we find in AS1063 is a new radio halo based on the lack of a clear optical counterpart, the central location, and the large physical extent.  From the flux density measurement in the L-band, we  calculate a monochromatic radio halo power of $P_{\rm 1.4GHz} = ( 2.63~\pm~0.18)~\times~10^{24}$ W~Hz$^{-1}$ using the equation
\begin{equation}
P_{\rm 1.4GHz} = 4\pi D_{\rm L}^{2} S_{\rm 1.4 GHz}(1+z)^{-(\alpha +1)}, 
\end{equation}
where $D_{\rm L}$ is the luminosity distance. Here, we adopted  $\alpha = -1.14$, and derived the flux density at 1.4~GHz $S_{\rm 1.4 GHz}$ from $S_{\rm 1.5 GHz}$ (scaling with the mentioned spectral index). 



\begin{figure}
   \centering
    \includegraphics[width=0.49\textwidth]{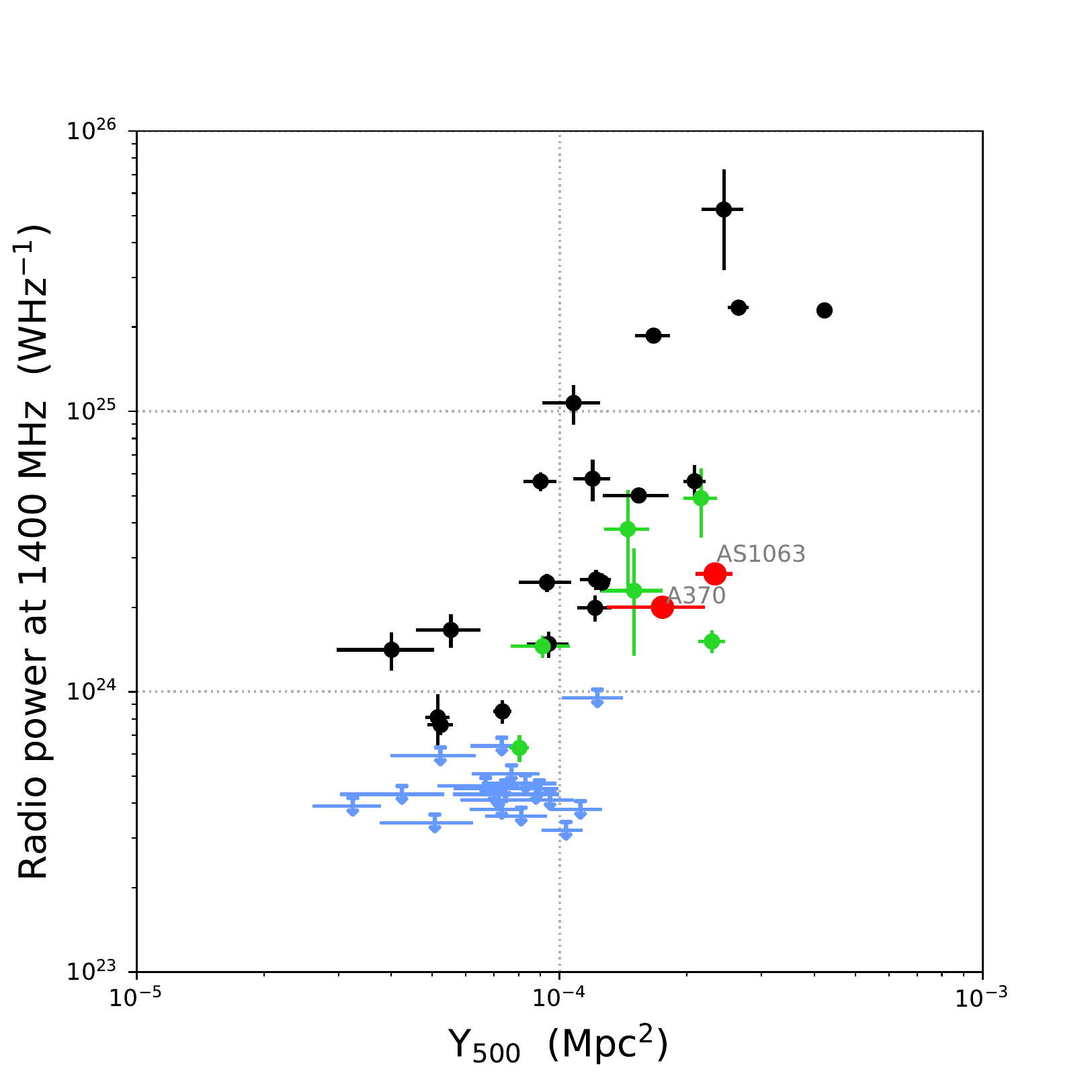}
      \caption{Distribution of radio halos in the $P_{\rm 1.4~GHz}$ - $Y_{\rm 500}$ diagram. Different symbols are used to represent the radio halos (black dots), radio halos with ultra-steep spectra (green dots) and the upper limits (blue arrows) from \cite{Cassano2013}.  AS1063 and A370 are shown as red dots.}
         \label{relations}
   \end{figure}

Statistical studies of radio halos have revealed a correlation between radio halo power and cluster mass \citep[e.g.,][]{2006MNRAS.369.1577C, Basu2012, Cassano2013,2014MNRAS.437.2163S,Martinez_Aviles2016}. The general idea is that a small fraction of the gravitational energy is converted into relativistic electrons during a cluster merger. The amount of energy released correlates with the cluster mass and results in the relation between radio power and cluster mass. Adopting the cluster sample from \cite{Cassano2013}, we over-plot our radio measurement of the halo emission on the $P_{\rm 1.4~GHz}$ - $Y_{\rm 500}$ diagram in Figure~\ref{relations}. $Y_{\rm 500}$ is the integrated SZ signal, {estimated from Y$_{5R_{500}}$ as Y$_{500}$ = Y$_{5R_{500}}$ / 1.79 \citep{Arnaud2010}. The Y$_{5R_{500}}$ is obtained from \cite{planck15} }. Unlike the X-ray luminosity, $Y_{\rm 500}$ is a more robust mass tracer as it less affected by the cluster's dynamical state \citep{Motl2005,Nagai2006}. Figure~\ref{relations} shows that the radio power of AS1063 is slightly lower than most clusters in this $Y_{\rm 500}$ range, but consistent within the expected scatter.


\begin{table*}
\caption{The properties of the radio sources for Abell S1063 and Abell 370, labelled in Figure~\ref{Lband_IR_as1063} and Figure~\ref{325_IR_a370}}             
\label{flux}      
\centering                          
\begin{tabular}{c c c c c c c c}        
\hline\hline                 
Name & RA & Dec & $S_{\rm 325~MHz}$ & $S_{\rm 1.5~GHz}$ & $S_{\rm 3~GHz}$ & $\alpha^{1500}_{325}$ & $\alpha^{3000}_{1500}$  \\    
  & (J2000) & (J2000) & (mJy) & (mJy) & (mJy) &   &  \\
\hline                        
AS1063 Halo$^{(a)}$ & 22 48 43.5 & -44 31 44.0 & 24.3 $\pm$ 2.5  & 5.8 $\pm$ 0.4 & 1.7 $\pm$ 0.2 &  -0.94 $\pm$ 0.08  & -1.77 $\pm$ 0.20 \\
AS1063   Source A & 22 48 44.0 & -44 31 51.85 & 8.5 $\pm$ 1.1  & 2.1 $\pm$ 0.2 & 1.0 $\pm$ 0.1 &  -0.93 $\pm$ 0.10 & -1.03 $\pm$ 0.15 \\
AS1063   Source B & 22 48 41.8 & -44 31 56.48 & 2.2 $\pm$ 0.3  & 0.7 $\pm$ 0.1 & 0.3 $\pm$ 0.1 &  -0.80 $\pm$ 0.12 & -1.07 $\pm$ 0.23\\  
AS1063   Source C & 22 48 44.6 & -44 30 09.59 & 40.7 $\pm$ 4.2 & 10.0 $\pm$ 0.5 & 4.5 $\pm$ 0.2  &  -0.92 $\pm$ 0.08 & -1.16 $\pm$ 0.10\\
AS1063   Source D & 22 48 49.3 & -44 30 44.58 & 37.8 $\pm$ 3.8  & 5.6 $\pm$ 0.3 & 1.8 $\pm$ 0.1 &  -1.24 $\pm$ 0.07 & -1.61 $\pm$ 0.11 \\
\hline

A370   Halo$^{(b)}$ & 02 39 52.0 & -01 35 12.0 & $20.0 \pm 2.3$  & $3.7 \pm 0.3$ & \textless~1.3 &  $-1.10 \pm 0.09$  & \textless~$-1.51$ \\
A370   Source A & 02 39 52.7 & -01 34 19.8 & $0.9 \pm 0.1  $  & $0.2 \pm 0.1$ & -- &  $-1.06 \pm 0.18$  &-- \\   
A370   Source B & 02 39 53.1 & -01 34 56.0 & --  & 0.11 $\pm$ 0.01$^{c}$ & -- & --  &-- \\   
A370   Source C & 02 39 50.9 & -01 35 42.4 & $3.1 \pm 0.3$  & $1.5 \pm 0.1$ & $0.8 \pm 0.1$ &  $-0.48 \pm 0.09$  &$-0.91 \pm 0.16$ \\ 
A370   Source D & 02 39 52.1 & -01 35 56.8 & $1.2 \pm 0.6$  & $0.3 \pm 0.1 $ & $0.2 \pm 0.1$ &  $-0.86 \pm 0.34$  &$-0.91 \pm 0.75$ \\   
A370   Source E & 02 39 55.4 & -01 34 07.9 & $15.5 \pm 2.0$  & $7.1 \pm 0.7$ & $ 4.4 \pm 0.4$ &  $-0.51 \pm 0.10$  &$-0.70 \pm 0.19$ \\  
A370   Source F & 02 39 56.5 & -01 34 29.0 & $4.5 \pm 1.0$  & $2.0 \pm 0.4$ & $1.3 \pm 0.1$ &  $-0.53 \pm 0.19$  &$-0.64 \pm 0.28$ \\
\hline

\end{tabular}
\tablefoot{$^{(a, b)}$ Alternative flux densities by manually subtraction the contribution of compact sources. For AS1063: $S_{\rm 325~MHz}= 25.3 \pm 2.7$ mJy, $S_{\rm 1.5~GHz}=5.1 \pm 0.5$ mJy, $S_{\rm 3~GHz}=1.4 \pm 0.2$ mJy; For A370: $S_{\rm 325~MHz}= 23.9 \pm 2.9 $ mJy, $S_{\rm 1.5~GHz}= 2.7 \pm 0.8 $ mJy, $S_{\rm 3~GHz} \textless 1.7 $ mJy.\\
$^{(c)}$~The southern BCG is blended with other sources. Here we adopt the 1.4~GHz flux density from \cite{Wold2012}.}

\end{table*}

\subsubsection{Spectral index maps}



\begin{figure*}
   \centering
    \includegraphics[width=1.0\textwidth]{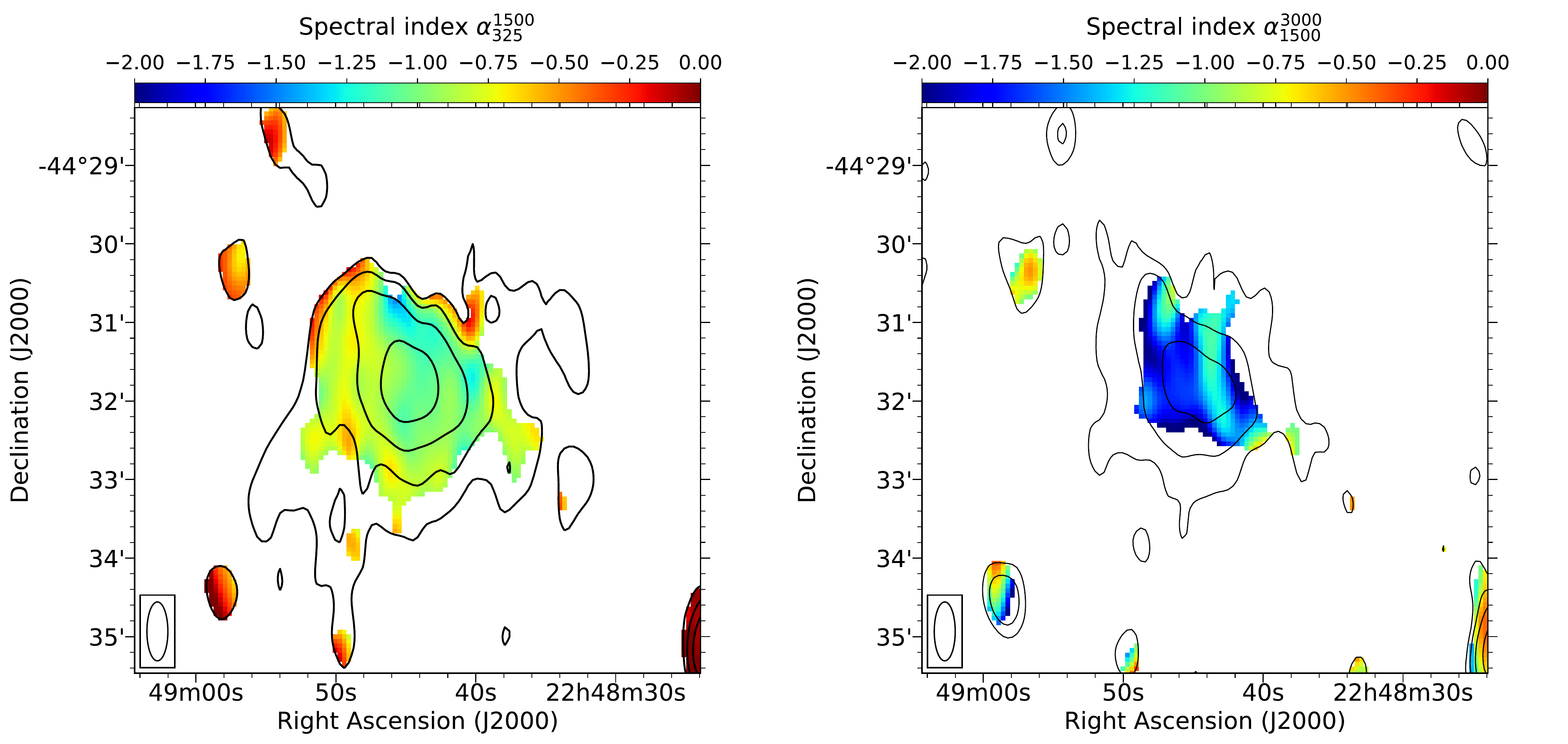}
      \caption{Spectral index maps of AS1063 between 325~MHz and 1.5~GHz (left), and between 1.5~GHz and 3.0~GHz (right). The radio contours are from the 325~MHz (left) and 1.5~GHz (right) images. These maps were created from the compact source subtracted uv-data. The beam size is indicated in the bottom left corner. Contour levels are drawn at [1, 2, 4, 8, ...] $\times$ 3$\sigma_{\rm rms}$, where $\sigma_{\rm 325~MHz}$= 96 $\mu$Jy~beam$^{-1}$, $\sigma_{\rm 1.5~GHz}$= 40 $\mu$Jy~beam$^{-1}$.}
         \label{spec_index_map}
   \end{figure*}

We constructed spectral index maps between 0.325 and 1.5~GHz, and between 1.5 and 3.0~GHz to characterize the spectral index distribution across the halo of AS1063 with resolution of $45\arcsec \times 16\arcsec$, see Figure~\ref{spec_index_map}. The imaging details can be found in Section~2.4. Spectral index uncertainty maps can be found in Appendix~\ref{sec:spixerror}.

The spectral index between 0.325 and 1.5~GHz ranges from $\sim$-0.8 at the eastern part of the halo to $\sim$-1.1 at the western part. The halo has steeper values between 1.5~GHz and 3.0~GHz, varying from $\sim$-1.7 at the eastern side to $\sim$-1.25 at the western side. {However, these east-west trends are not very significant, considering the spectral index uncertainty of $\sim$0.15 and $\sim$0.25 in $\alpha^{1500}_{325}$ and $\alpha^{3000}_{1500}$ maps, respectively.}







\begin{figure}
   \centering
    \includegraphics[width=0.49\textwidth]{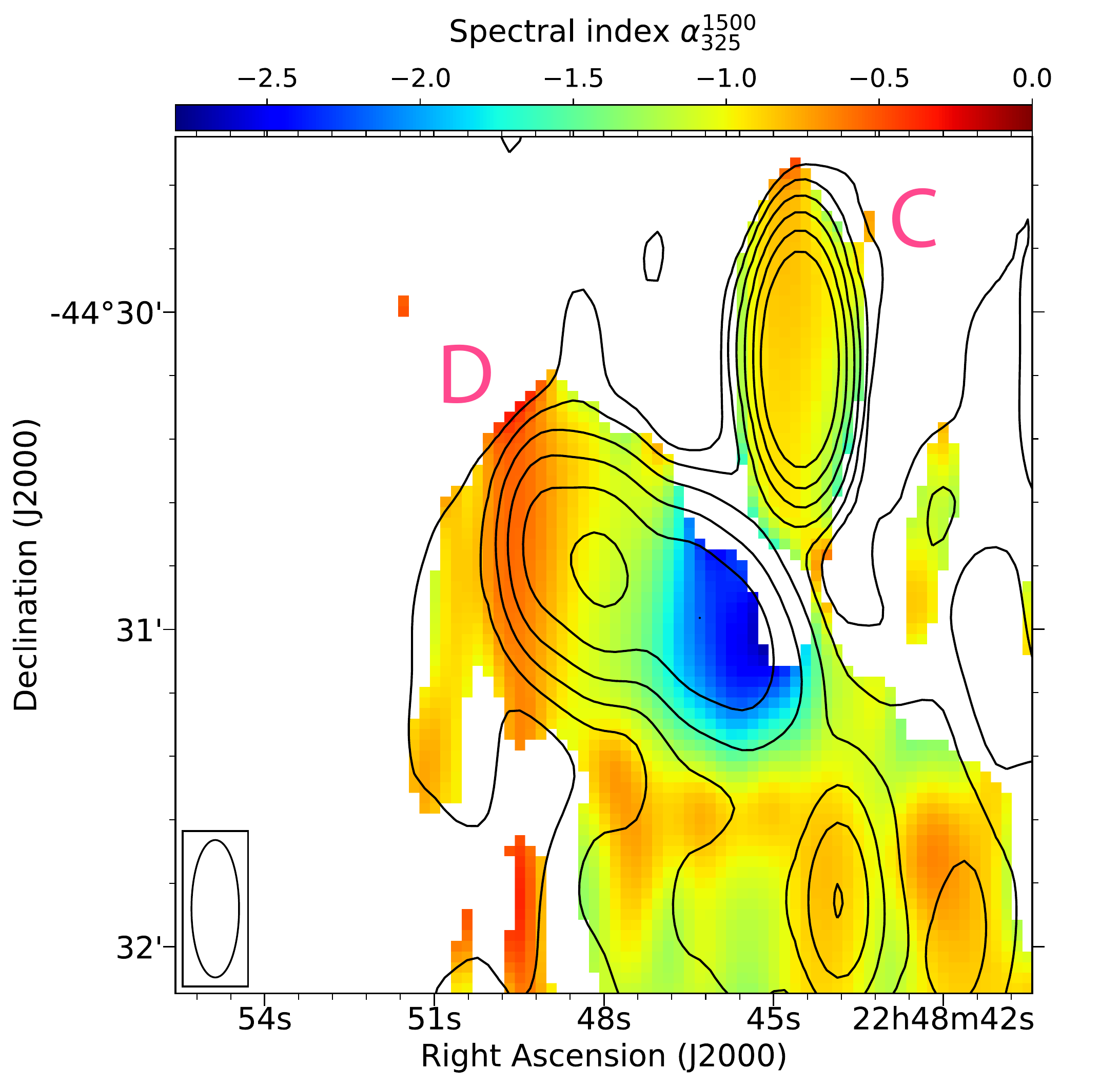}
      \caption{Spectral index map of source D in AS1063 between 325~MHz and 1.5~GHz. The radio contours are  from the GMRT 325~MHz image.  The beam size is 26.0\arcsec $\times$ 9.0\arcsec and indicated in the bottom left corner. Contour levels are drawn at [1, 2, 4, 8, 16, 32 ...] $\times$ 5$\sigma_{\rm rms}$, where $\sigma_{\rm rms}$ = 46 $\mu$Jy~beam$^{-1}$.}
         \label{spec_index_map_SourceD}
   \end{figure}

Figure~\ref{spec_index_map_SourceD} shows the spectral index map of source~D between 325~MHz and 1.5~GHz. The corresponding uncertainty map can be found in Figure~\ref{error_spec_index_ma_SourceD}. 
Source~D shows a clear spectral index gradient along the radio tail. The spectral index steepens from $\sim-0.6$ at location of the IR counterpart to $\sim-2.4$ at the tail end. Such a trend is expected as the result of the radiative losses of electrons. 

   \begin{figure}
   \centering
    \includegraphics[width=0.49\textwidth]{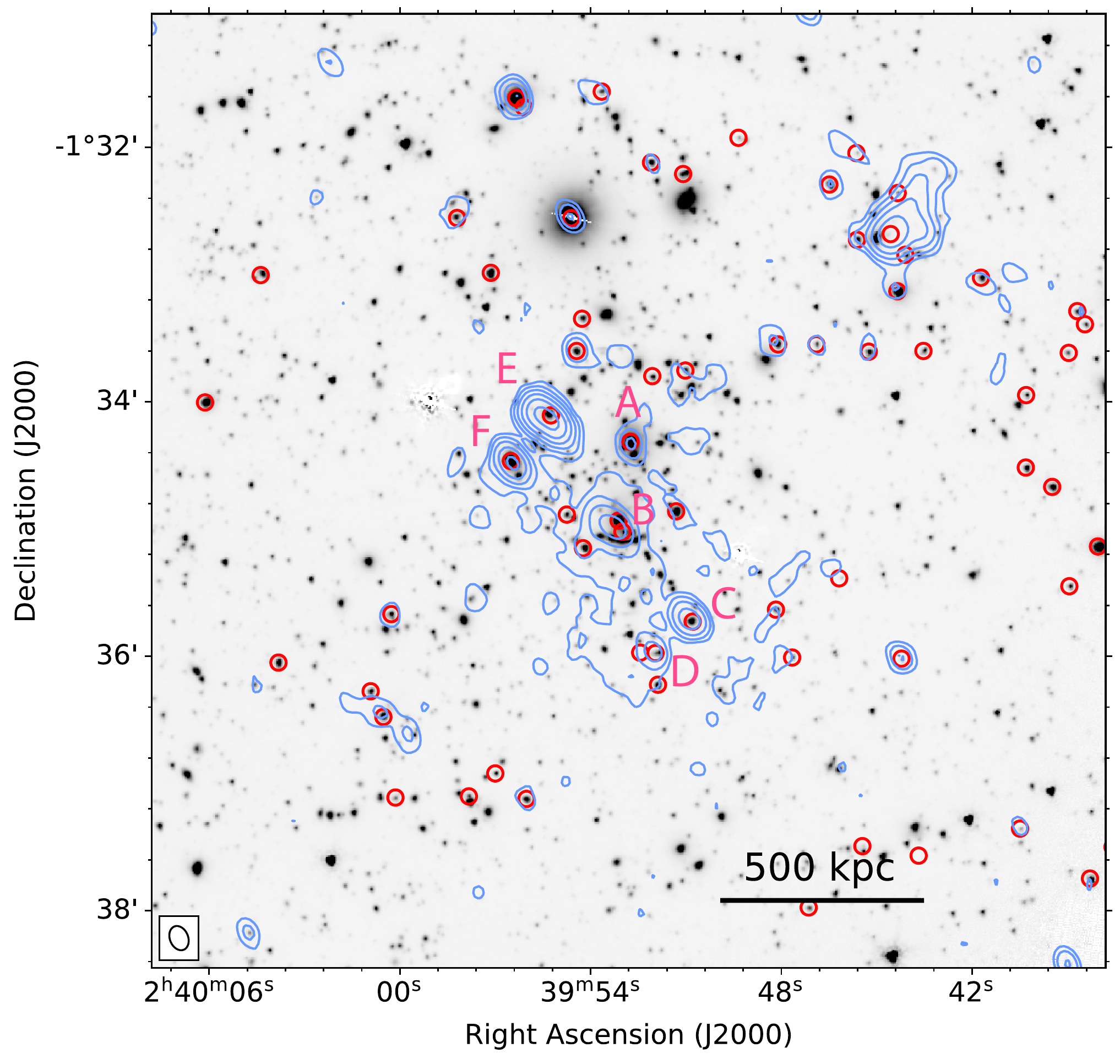}
      \caption{{\sl Spitzer} 3.6~$\mu$m IRAC image of A370 overlaid with the GMRT 325~MHz radio contours. The beam size is 11.9\arcsec $\times$ 8.6\arcsec. Contour levels are drawn at [1, 2, 4, 8, 16, 32] $\times$ 3$\sigma_{\rm rms}$, where $\sigma_{\rm rms}$ = 62 $\mu$Jy~beam$^{-1}$. Compact radio sources are indicated with red labels from A to F. The red circles mark the 1.4~GHz radio source catalog from \cite{Wold2012}. }
         \label{325_IR_a370}
   \end{figure}

   \begin{figure*}
   \centering
    \includegraphics[width=0.99\textwidth]{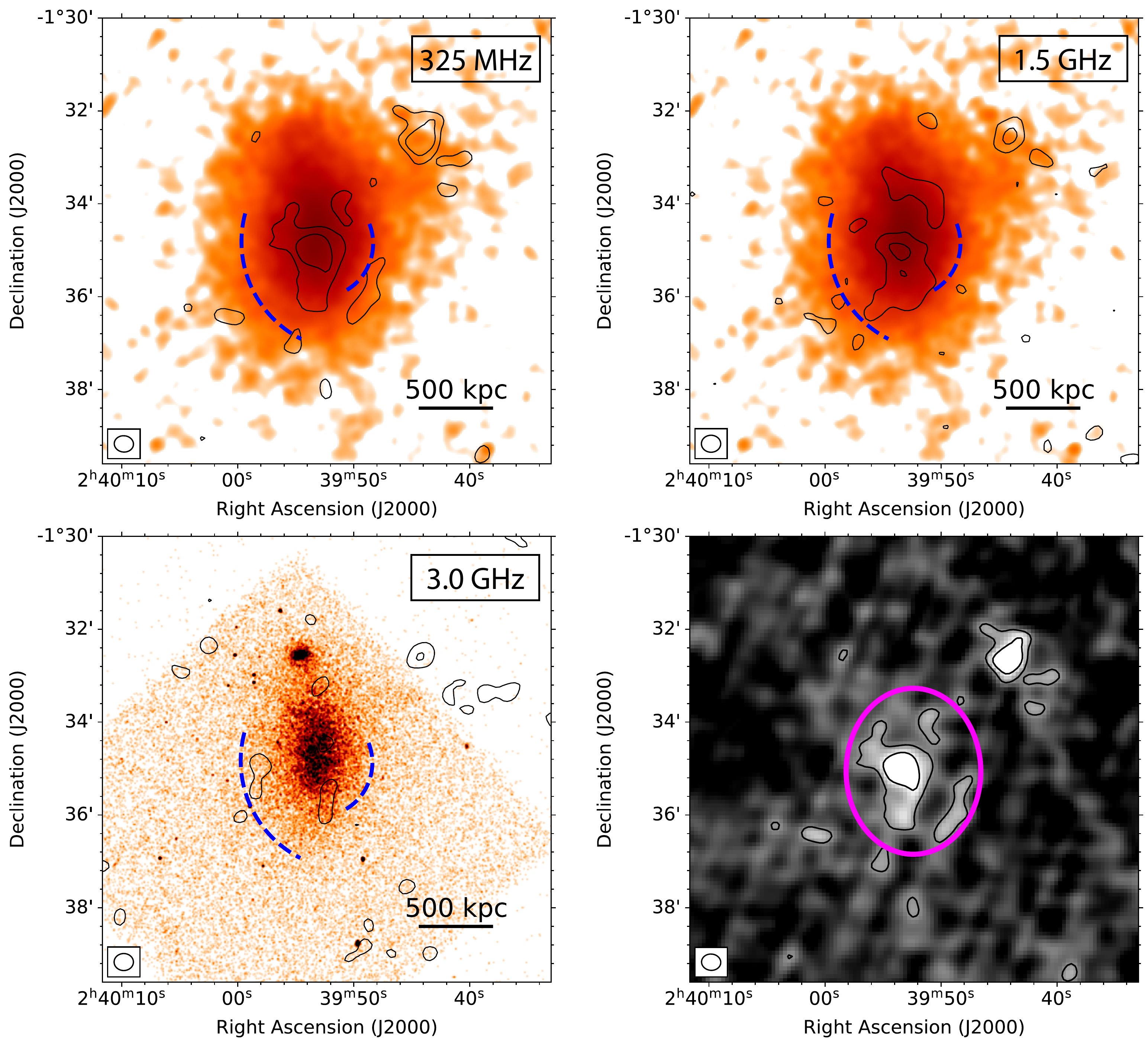}
      \caption{\xmm 0.3--7~keV image of A370 overlaid with the radio contours with compact sources subtracted from the uv-data (325~MHz, top left; 1.5~GHz, top right). 
      All the point source in \xmm image are removed and smoothed with a Gaussian with scale of 3 pixels across. The Blue dashed curves show the surface brightness edges found by \cite{Botteon2018}. Bottom left: \chandra 0.5--2.0~keV image overlaid with the 3.0~GHz radio contours with compact sources subtracted from the uv-data.  
      The radio beam size is 25.0\arcsec $\times$ 22.0\arcsec. Contour levels are drawn at [1, 2, 4, 8, ...] $\times$ 3$\sigma_{\rm rms}$, where $\sigma_{\rm 325~MHz}$= 161 $\mu$Jy~beam$^{-1}$, $\sigma_{\rm 1.5~GHz}$= 34 $\mu$Jy~beam$^{-1}$, $\sigma_{\rm 3.0~GHz}$ = 18 $\mu$Jy~beam$^{-1}$. The inner uv-cut of 0.22~k$\lambda$ is adopted for 325~MHz and 1.5~GHz images. Bottom right: GMRT 325~MHz image of A370 depicting the region (magenta ellipse) where we extract the integrated flux densities.}
         \label{radio_X-ray_a370}
   \end{figure*}

   \begin{figure}
   \centering
    \includegraphics[width=0.51\textwidth]{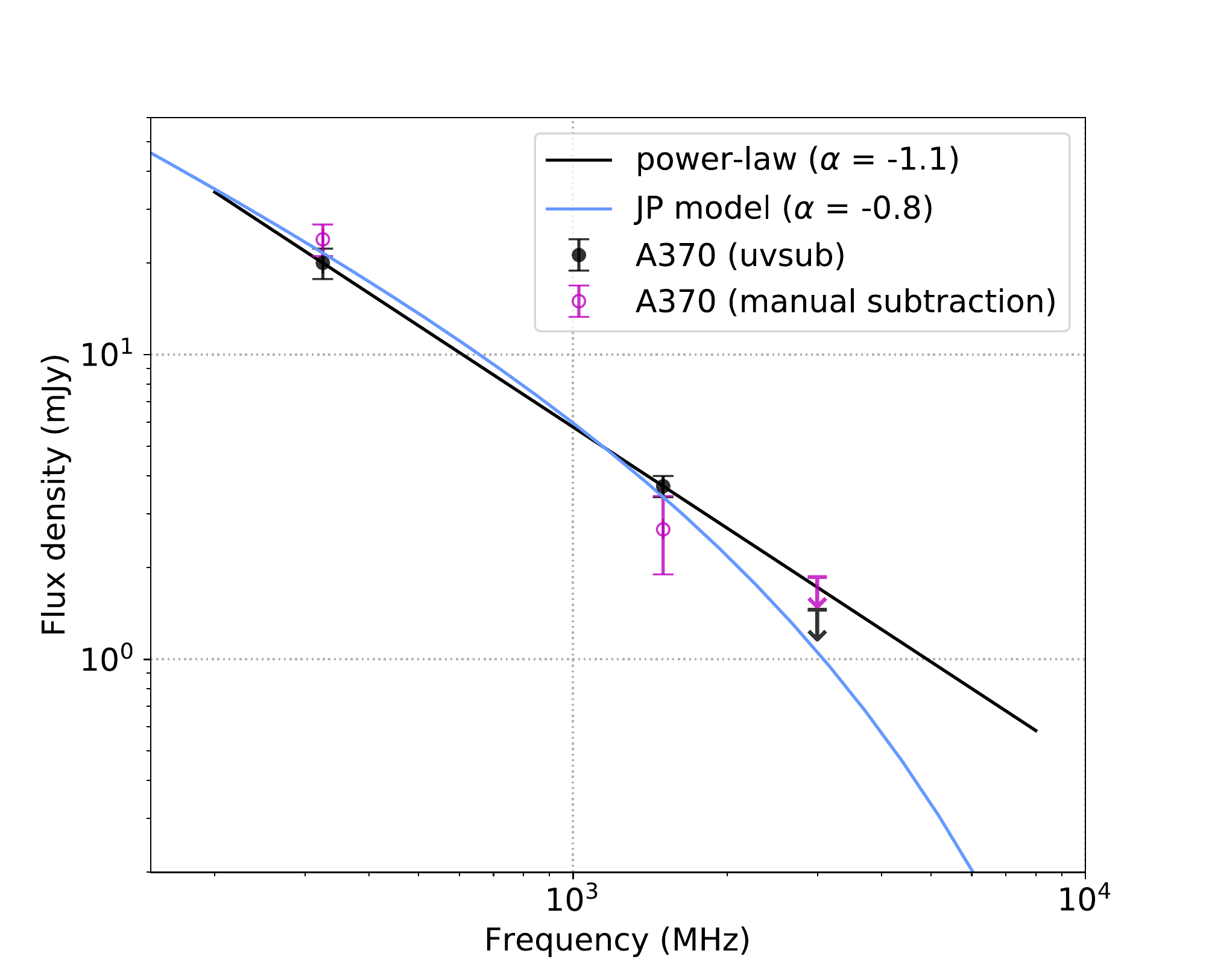}
      \caption{Integrated flux densities and 3$\sigma$ upper-limits for the halo in A370 at 325~MHz, 1.5~GHz, and 3.0~GHz, measured by two methods. The black line shows a single power-law with spectral index $\alpha = -1.1$. The JP model has an injection spectral index of $-0.8$.}
         \label{spec_index_a370}
   \end{figure}

\subsection{Abell\,370}
A {\sl Spitzer} image of A370 with 325~MHz radio contours overlaid is presented in Figure~\ref{325_IR_a370}.  A number of compact radio sources are detected in the cluster field.  
\cite{Wold2012} analyzed 1.4~GHz VLA observations of the A370 cluster field in the A and B configurations. Their radio source catalog is over plotted on Figure~\ref{325_IR_a370}. We also label several radio sources in the cluster region, source A to F. The flux densities of these sources are reported in Table~\ref{flux}. 



There are two BCGs in A370 and both show radio emission associated with an AGN. Source~A, the northern BCG is detected at 325~MHz and 1.5~GHz. The flux densities are $S_{\rm 325~MHz}$ = \textbf{$0.86 \pm 0.10$}~mJy and $S_{\rm 1.5~GHz}$ = \textbf{$0.17 \pm 0.04$}~mJy, corresponding to a spectral index of \textbf{$-1.06 \pm 0.18$}. The southern BCG marked as B, blends with other nearby radio sources in our relatively low resolution image with a beam size of 11.9\arcsec $\times$ 8.6\arcsec. The high resolution 1.4~GHz image of \cite{Wold2012} shows a separate radio source associated with the BCG with a flux density of 0.11~mJy. 



Some very faint extended emission is found in the southern part of the cluster near source B and south of it (see  Figure~\ref{325_IR_a370}). 
To better bring out the diffuse emission we made low-resolution compact source subtracted images (for details see Section~\ref{sec:sourcesubtraction}). A \xmm X-ray image, with radio contours at 0.325 and 1.5~GHz overlaid, is displayed in Figure~\ref{radio_X-ray_a370}. For completeness, a \chandra image is also shown, overlaid with radio contours at 3~GHz.
Diffuse emission is detected at 325~MHz and 1.5~GHz. The total extent of the diffuse emission is about 500--700~kpc. The surface brightness of this emission peaks at the location of the southern BCG. 
Note that the emission at north-west periphery of the cluster are residuals from an extended tailed radio galaxy that is not completely subtracted from uv-data. The flux density of the radio halo is $20.0 \pm 2.3$~mJy at 325 MHz and $3.7 \pm 0.3$~mJy at 1.5~GHz. For the non-detection at 3.0~GHz we determine a $3\sigma$ upper limit of \textless~1.3~mJy. The region where we measured the flux densities is indicated in Figure~\ref{radio_X-ray_a370}. The corresponding spectral indices are $\alpha^{1500}_{325} = -1.10~\pm~0.09$, and $\alpha^{3000}_{1500} < -1.51 $. 

We classify the extended emission in A370 as a radio halo based on the lack of a clear optical counterpart, physical extent, and location in the cluster. Deeper observations will be needed to determine the full extent of the radio halo given its low surface brightness. We compute a monochromatic radio power at 1.4~GHz of $P_{\rm 1.4GHz} = (2.00~\pm~0.16)~\times~10^{24}$ W~Hz$^{-1}$, where we take $\alpha = -1.1$. The radio halo power is consistent with that expected from the known scaling relations \citep{Cassano2013}, see Figure~\ref{relations}. We do not compute spectral index maps of the radio halo as the emission is barely detected at three times the map noise level.

{\cite{Botteon2018} detected two X-ray surface brightness edges on the west and east side of the cluster, shown as blue dashed curves in Figure~\ref{radio_X-ray_a370}. Due to the limited number of X-ray counts, the nature (cold front or shock) of these edges could not be determined. No clear correspondence between the edges and radio emission halo emission is found.}

\section{Discussion}
\label{sec:discussion}



\subsection{Merger scenarios and nature of the radio halos}
The presence of a radio halo in the massive merging cluster A370 is expected based on the dynamical state of this cluster. A370 has a clear bimodal mass distribution, indicating a major merger event \citep{Richard2010}. This is also supported by the presence of X-ray surface brightness edges \citep{ Botteon2018}. The radio halo in A370 thus supports the general scenario that radio halos trace particles that are re-accelerated by merger induced turbulence.

The dynamical state of AS1063 is key to interpreting the nature and formation scenario for its radio halo. However, the dynamical state of AS1063 is still under debate. The presence of a single BCG and a morphological analysis from \xmm observations suggest a relaxed dynamical state \citep{Lovisari2017}. This X-ray analysis used the concentration value, power ratio, and centroid shift. On the other hand the offset between the galaxy distribution and  the peak of the X-ray emission, high global X-ray temperature, and weak lensing analysis indicate an ongoing major merger event \citep{Gomez2012, Gruen2013}. 

\begin{figure}
   \centering
    \includegraphics[width=0.49\textwidth]{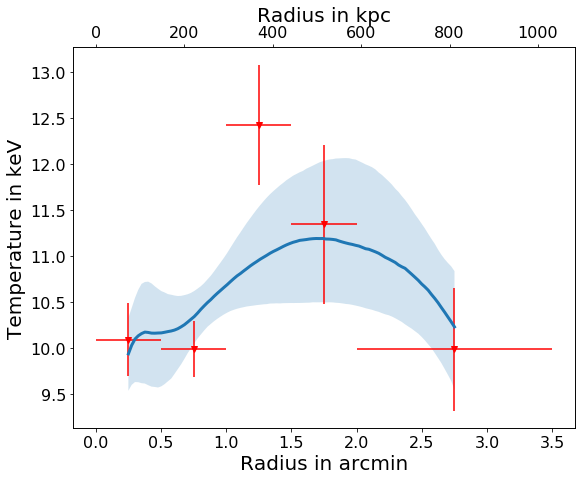}
      \caption{The radial temperature profile of AS1063 obtained from \xmm observations (Lovisari et al. in prep.). The size of the bins was set to be $\geq$ 30\arcsec~to limit the scatter of photons from bin to bin. The observed projected temperatures are shown by red points. The blue line and shaded area show the best-fit projected profile and its 1$\sigma$ uncertainty computed by using the best-fit three-dimensional model described in \cite{2006ApJ...640..691V}.}
         \label{as1063_kt}
   \end{figure}

Radio mini-halos exclusively occur in cool core clusters. In addition, typically their sizes are a few hundreds of kpc and the emission from the mini-halo is confined to the cooling region \citep[e.g.,][]{2008ApJ...675L...9M, 2017ApJ...841...71G, Giacintucci2019}. 
To determine whether AS1063 hosts a cool core we derived a radial temperature profile from the 
\xmm data, see Sect.~\ref{sec:xray} for more details.
This radial profile is displayed in Figure~\ref{as1063_kt}. The central two bins of this profile (\textless~250~kpc) show temperatures about 10 keV which indicate AS1063 has a rather hot-core. This result is consistent with the conclusion of the presence of a hot-core based on \chandra data by \cite{Gomez2012}.  We thus conclude that the diffuse emission in AS1063 cannot be classified as a radio mini-halo, also in line with the rather large extent of the halo.
The presence of the giant radio halo ($\sim$1.2~Mpc) in AS1063 therefore suggests a link to a cluster merger event. In line with the previous claims of an on-going merger by \cite{Gomez2012, Gruen2013}.




\subsection{Curved radio spectra}
 
The radio spectra of halos provide important information about the underlying particle (re-)acceleration mechanisms. According to the turbulent re-acceleration model, we expect a cutoff in the energy spectrum of the CR electrons which also leads to a cutoff in the synchrotron emission above a certain frequency \citep[e.g.,][]{Brunetti2014}. However, only a few clusters have been found that show this spectral steepening. Some examples are the Coma Cluster \citep{Schlickeiser1987,Thierbach2003} and Abell\,3562 \citep{A3562}. 
A possible explanation for why very few clusters show spectral curvature are the observational difficulties involved. High quality flux density measurements at wide enough frequency spacing are hard to obtain. In addition, the  magnetic field properties and amount of turbulence will differ with location in the ICM. These inhomogeneous conditions result in spectra with different local cutoff frequencies. The spectral curvature will therefore be less pronounced for the integrated spectra when these inhomogeneous conditions are present \cite[e.g.,][]{2013ApJ...762...78Z, Donnert2013,Pinzke2017}.

Our flux density measurements of AS1063 show spectral steepening between 1.5 and 3~GHz, providing support for the turbulent re-acceleration model. To characterize the spectral steepening, we compare our flux measurements to a \cite{Jaffe1973} (JP) model with a ``reasonable'' injection spectral index of $-0.8$ \citep{Thierbach2003} in Figures~\ref{spec_index_as1063} and~\ref{spec_index_a370}. A comparison with a  JP model is useful here since a JP spectrum has a power-law shape with an exponential cutoff. The amount of spectral steepening for AS1063 is quite strong and well described by exponential cutoff. This would indicate that ICM conditions in AS1063 are rather homogeneous leading to similar cutoff frequencies at different spatial positions. For A370, the measurement uncertainties do not allow us to draw firm conclusions on the amount of spectral steepening.


It is important to stress that the observed spectral curvature for AS1063 is completely based on one measurement at 3~GHz. Therefore, future observations are important to confirm this result and characterize the shape of the radio spectrum in more detail.  We note that the relatively small spatial extent (3\arcmin) and relatively high surface brightness of the AS1063 radio halo make it a promising target for future observations with the VLA at 4--8~GHz and at $\lesssim 1$~GHz with the uGMRT.

\section{Conclusions}

\label{sec:conclusions}
In this paper, we presented 325~MHz GMRT and  1--4~GHz VLA observations of the Frontier Fields clusters AS1063 and A370. The results are summarized below:

\begin{enumerate}

\item We discovered a giant $\sim$1.2~Mpc radio halo in AS1063. The radio halo roughly follows the X-ray emission from the thermal ICM. We determined a radio halo power of $P_{\rm 1.4GHz} = (2.63 \pm 0.18) \times 10^{24}$~W~Hz$^{-1}$.

\item The integrated spectral index of the AS1063 radio halo measures $-0.94 \pm 0.08$ between 0.325 and 1.5~GHz and it steepens to $-1.77 \pm 0.20$ between 1.5~GHz and 3.0~GHz. 
This spectral steepening provides support for the turbulent re-acceleration model for the formation of radio halos.

\item We discovered a faint radio halo in A370 with a size of about 500--700~kpc. The radio halo power is $P_{\rm 1.4GHz} = (2.00 \pm 0.16) \times 10^{24}$~W~Hz$^{-1}$. The radio halo is not detected at 3.0~GHz, with a 3$\sigma$ limit of 1.3~mJy. We measure a spectral index of $-1.10~\pm~0.09$ between 0.325 and 1.5~GHz.

\item The radio halo powers of AS1063 and A370 follow the $P_{\rm 1.4~GHz}$ - $Y_{\rm 500}$ scaling relation. Complementing our radio data with \chandra and \xmm X-ray observations provides support for the idea that both radio halos are related to ongoing cluster merger events.


\end{enumerate}

\begin{acknowledgements}

The National Radio Astronomy Observatory is a facility of the National Science Foundation operated under cooperative agreement by Associated Universities, Inc. We thank the staff of the GMRT that made these observations possible. GMRT is run by the National Centre for Radio Astrophysics of the Tata Institute of Fundamental Research. The scientific results reported in this article are based in part on observations made by data obtained from the Chandra Data Archive. Based on observations obtained with XMM-Newton, an ESA science mission with instruments and contributions directly funded by ESA Member States and NASA. 
RJvW and AB acknowledge support from the VIDI research programme with project number 639.042.729, which is financed by the Netherlands Organisation for Scientific Research (NWO). LL acknowledges support from NASA through contracts 80NSSCK0582 and 80NSSC19K0116. Basic research in radio astronomy at the Naval Research Laboratory is supported by 6.1 Base funding. This research made use of Astropy,\footnote{http://www.astropy.org} a community-developed core Python package for Astronomy \citep{astropy:2013, astropy:2018}. This research made use of APLpy, an open-source plotting package for Python \citep{aplpy2012, aplpy2019}.

\end{acknowledgements}

\begin{appendix}

\section{The spectral index uncertainty maps}
\label{sec:spixerror}

\begin{figure*}
   \centering
    \includegraphics[width=0.99\textwidth]{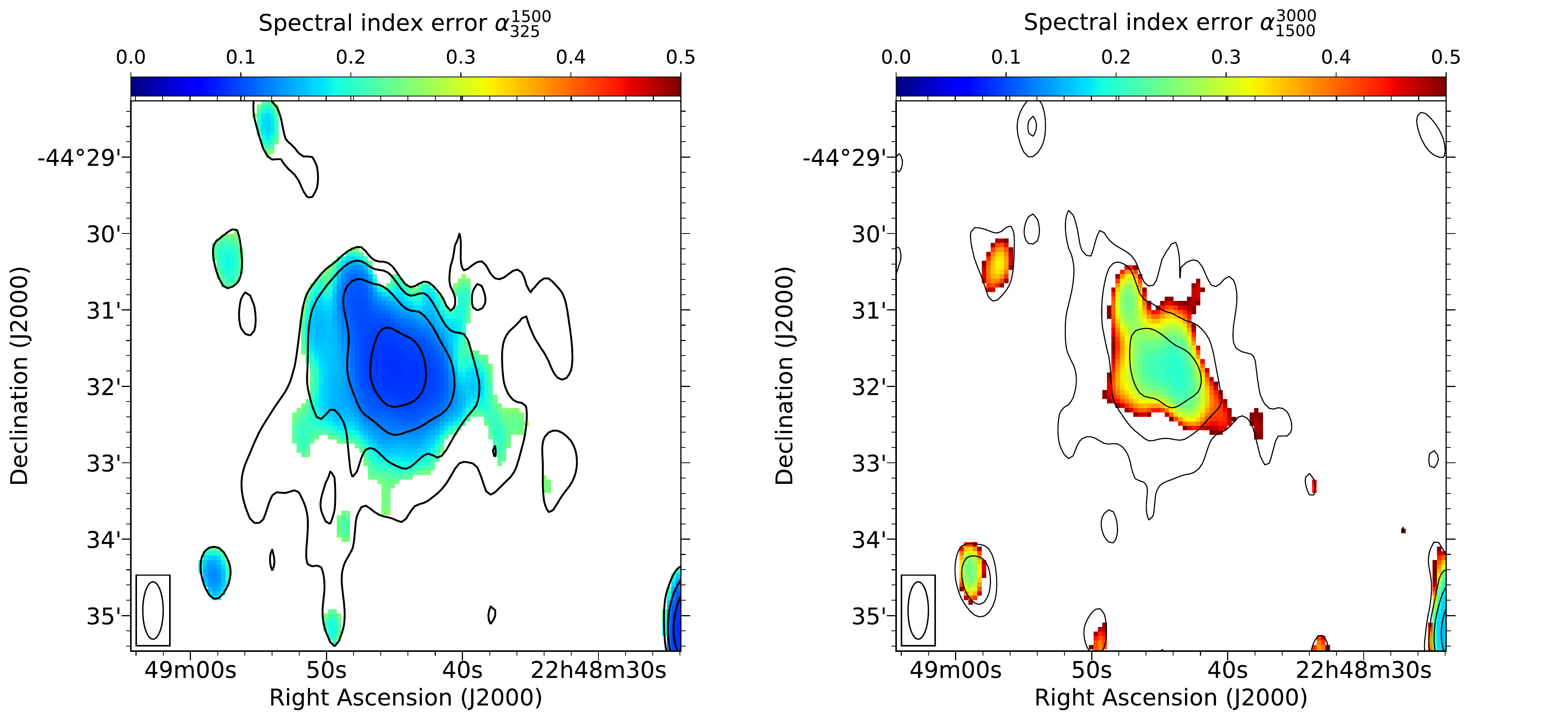}
      \caption{Spectral index uncertainty maps of AS1063 between 325~MHz and 1.5~GHz (left), and between 1.5 GHz and 3.0~GHz (right). The radio contours are from the 325~MHz (left) and 1.5~GHz (right) images.
              }
         \label{error_spec_index_map}
   \end{figure*}

\begin{figure}
   \centering
    \includegraphics[width=0.49\textwidth]{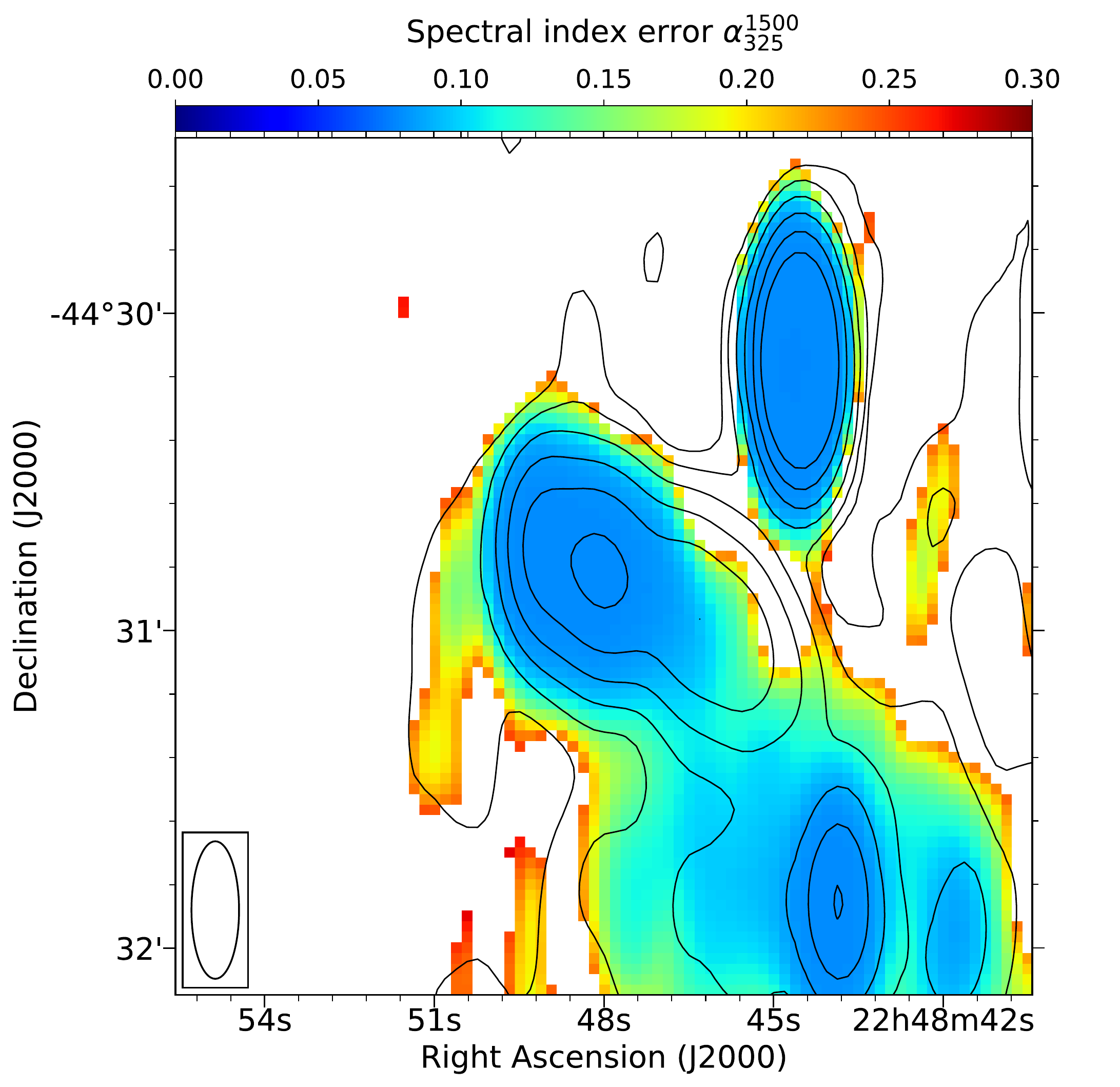}
      \caption{Spectral index uncertainty map of source D in AS1063 between 325~MHz and 1.5~GHz. The radio contours are from the 325~MHz image.
              }
         \label{error_spec_index_ma_SourceD}
   \end{figure}

Figures~\ref{error_spec_index_map}~and~\ref{error_spec_index_ma_SourceD} show the uncertainty maps corresponding to the spectral index maps in Figures~\ref{spec_index_map}~and~\ref{spec_index_map_SourceD}, respectively. 

\end{appendix}

\bibliographystyle{aa}
\bibliography{ms}

\end{document}